\pdfoutput=1

\documentclass[aps,prl,reprint,showpacs,superscriptaddress]{revtex4-1}

\usepackage{graphicx}
\usepackage{graphics}
\usepackage{amsmath}
\usepackage{amssymb}
\usepackage{amsfonts}
\usepackage{dcolumn}
\usepackage{dsfont}
\usepackage{latexsym}
\usepackage{rotating}
\usepackage{color}
\usepackage{latexsym}
\usepackage{bbm}
\usepackage{subfigure}
\usepackage{float}
\usepackage{epsfig}
\usepackage{epsf}
\usepackage{psfrag}
\usepackage{bm}
\usepackage{amsthm}
\usepackage{eucal}
\usepackage{mathrsfs}
\usepackage{url}
\usepackage{braket}
\usepackage{epstopdf}
\graphicspath{ {./images/} }
\newcommand{\RNum}[1]{\uppercase\expandafter{\romannumeral #1\relax}}
\usepackage{color} 

\usepackage{hyperref}
\hypersetup{
	colorlinks=true,final=true,
	linkcolor=blue,
	citecolor=blue,
	filecolor=blue,
	urlcolor=blue,
}

\begin{document}

	\title{In-plane polarization induced ferroelectrovalley coupling in a two-dimensional rare-earth halide}
	
	\author{Srishti Bhardwaj}
	\affiliation{ Department of Physics, Indian Institute of Technology Roorkee, Roorkee - 247667, Uttarakhand, India}
	\author{T. Maitra}
	\email{tulika.maitra@ph.iitr.ac.in}
	\affiliation{ Department of Physics, Indian Institute of Technology Roorkee, Roorkee - 247667, Uttarakhand, India}
	
	\date{\today}

\begin{abstract}
We propose a mechanism where the valley splitting is caused by an in-plane electric polarization and the coupling between the two makes it possible for an electric field to control the valley degree of freedom. We demonstrate this by considering Gd-substituted EuCl$_2$ monolayer in its 1T-phase using first-principles calculations. This monolayer exhibits an in-plane polarization which breaks the inversion symmetry of the monolayer leading to a spontaneous valley splitting. The resulting valley polarization is strongly coupled with the electric polarization and, hence, the valley degree of freedom can be switched by an external electric field in this case, instead of the conventional magnetic field. We show that a similar ferroelectric-ferrovalley (FE-FV) coupling can also exist in the previously reported ferroelectric (CrBr$_3$)$_2$Li monolayer. This mechanism opens up a new avenue for electric field control of valley polarization in two-dimensional materials.
\end{abstract}

	\maketitle

After charge and spin, valley is the new degree of freedom that is highly sought after in modern electronic devices \cite{1,2}. So far, the conventional method to control the valley degree of freedom has been via the spin-valley coupling  \cite{3,4}. However, valley switching using an external magnetic field is a highly power-consuming process and is not ideal for miniaturized devices. Therefore, just like the electric control of magnetism, an electric control of the valley degree of freedom is also an important requirement \cite{5}. This phenomenon can be realized via a coupling between ferroelectricity and valley degree of freedom so that when the electric polarization is reversed by an external electric field, valley degree of freedom is also switched accordingly.
	
It is difficult to find materials that have both reversible electric polarization and valley polarization, thus making the electric field control of valley polarization an elusive task. Valley polarization has been primarily found in materials like transition-metal dichalcogenides in 2H phase where inversion symmetry is inherently broken in their monolayers \cite{6}. However, one cannot have an in-plane 180$^\circ$ reversible polarization in such materials due to their $C_3$ rotational symmetry. Thus far, only a small number of materials with electric-field control of the valley have been predicted using first-principles calculations \cite{7,8,9,10,11}. Some of the intrinsic materials to possess this phenomenon are found to have an out-of-plane polarization coupled with the structure crystal field, which results in a coupling of the valley degree of freedom with the direction of electric polarization. This mechanism is found in materials with a special symmetry, such as a breathing kagome lattice as in Ti$_3$B$_8$ \cite{7} and CuCr$_2$PS$_6$\cite{8}. The monolayers, which have an in-plane electric polarization coupled with the valley, are the transition metal monochalcogenides \cite{9}. However, they are not magnetic and have a zero Berry curvature and hence cannot display anomalous valley Hall effect. The carbon nitrides adsorbed by metal halides MCl$_2$ (M = Zn, Mg), MCl$_2$(C$_2$N)$_6$ (M=Zn, Mg) have a valley polarization but are also non-magnetic \cite{10}. Further, the polarization direction in these systems, can only be rotated by 120$^\circ$ and cannot be reversed by 180$^\circ$ due to the $C_3$ rotational symmetry of carbon nitrides. The ferroelectric-valley coupling has also been found in bilayers stacked at various angles \cite{11}. It has been shown, however, that in antiferromagnetic-ferroelectric bilayer 2H-VSe$_2$, a reversal of polarization is coupled with a switching of magnetisation as well, thus bringing the band structure back to its initial form. Therefore, ferromagnetism is coupled with ferroelectricity but a valley switching is not possible by an electric field alone. All these ferroelectric-valley coupled materials discussed above are transition-metal based, which usually have a smaller magnetic signal and magnetic anisotropy, whereas, the rare-earth metal-based materials possess large magnetic anisotropy and provide larger magnetic signal because of 4f-electrons \cite{12}. 
	
Recently, a number of rare-earth halides based multiferroics have been predicted, including ferroelastic (FA), ferroelectric (FE), ferromagnetic (FM), and ferrovalley (FV) orderings \cite{12,13,14,15,16,17,18,19}. However, the coupling between ferroelectricity and ferrovalley ordering has not yet been found in these materials. Here, we present a mechanism where the valley splitting is caused by an induced in-plane polarization in a system where the inversion symmetry is not inherently broken. Here, the spin-orbit coupling (SOC) associated with the in-plane polarization breaks valley degeneracy for the time-reversal symmetry-connected {\bf k}-points. Thus, by inverting the direction of polarization, the SOC also changes its sign, causing the switching of the valley degree of freedom. To realize this mechanism, we introduce a 1/3rd Gd-substitution in 1T-EuCl$_2$, which initially has a $C_6$ symmetry and thus, can carry a reversible electric polarization. The system undergoes a charge ordering due to the substitution that breaks its inversion symmetry and gives rise to an in-plane electric polarization. Ultimately, we get a FE-FM-FV material in which the valley degree of freedom is coupled with the FE polarization and can be controlled by an external electric field. Here, the inversion symmetry breaks as a result of polarization, in contrast to the conventional ferrovalley materials, where the inversion symmetry is inherently broken. The valley polarization, observed here, is a secondary effect and appears as a response to electric polarization. Thus, it gives a different/new route to create valley-polarized materials via ferroelectricity.
	\begin{figure}[!t]
		\centering
		\includegraphics[width=9.2 cm]{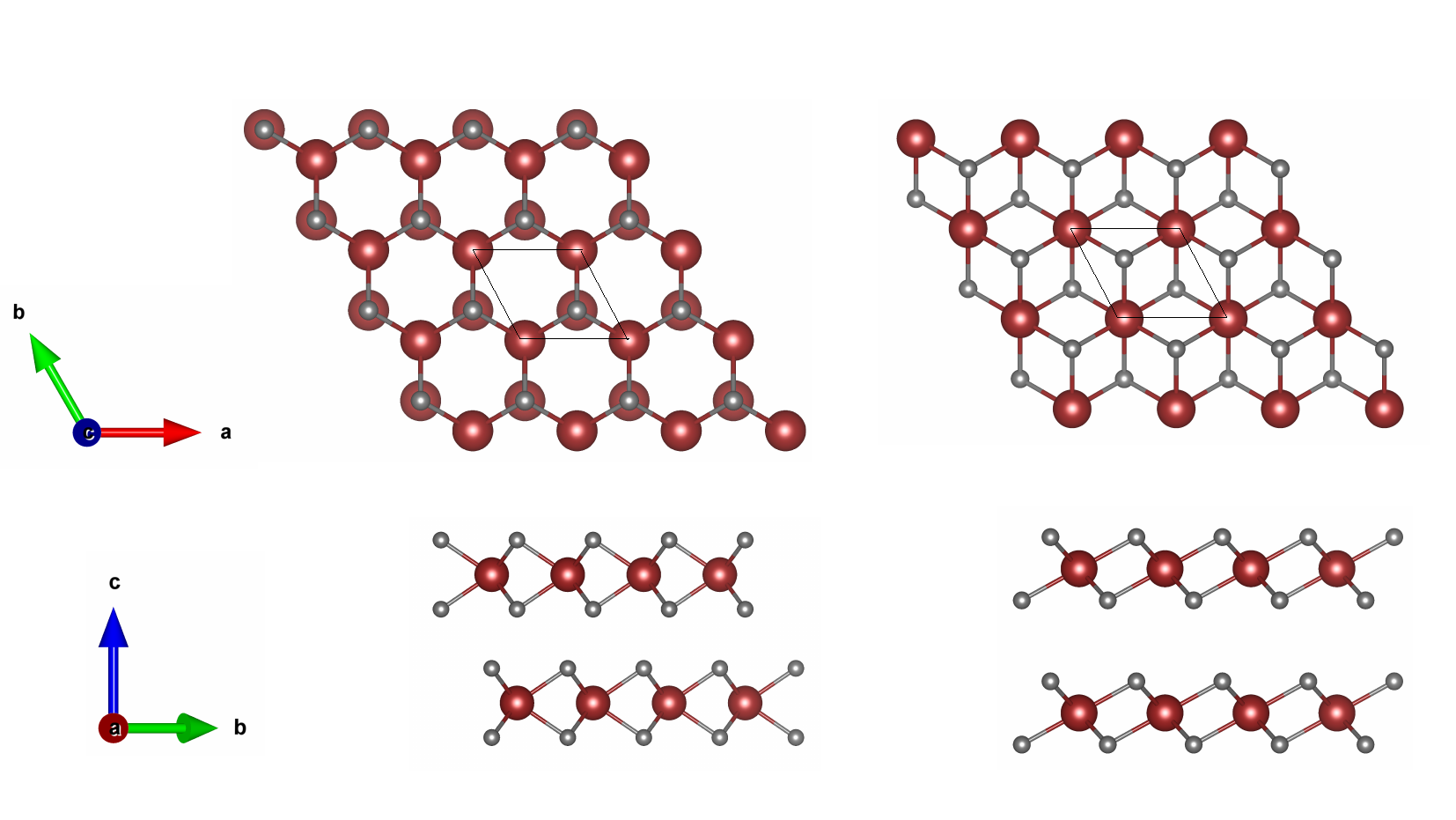}
		\caption{(a) 2H-phase of EuCl$_2$ (b) 1T-phase of EuCl$_2$ bulk (Brown spheres denote Eu-atoms, silver denote Cl atoms).}
	\end{figure}
	
	%

	\begin{figure*}[!t]
		\centering
		\includegraphics[width=18.0cm]{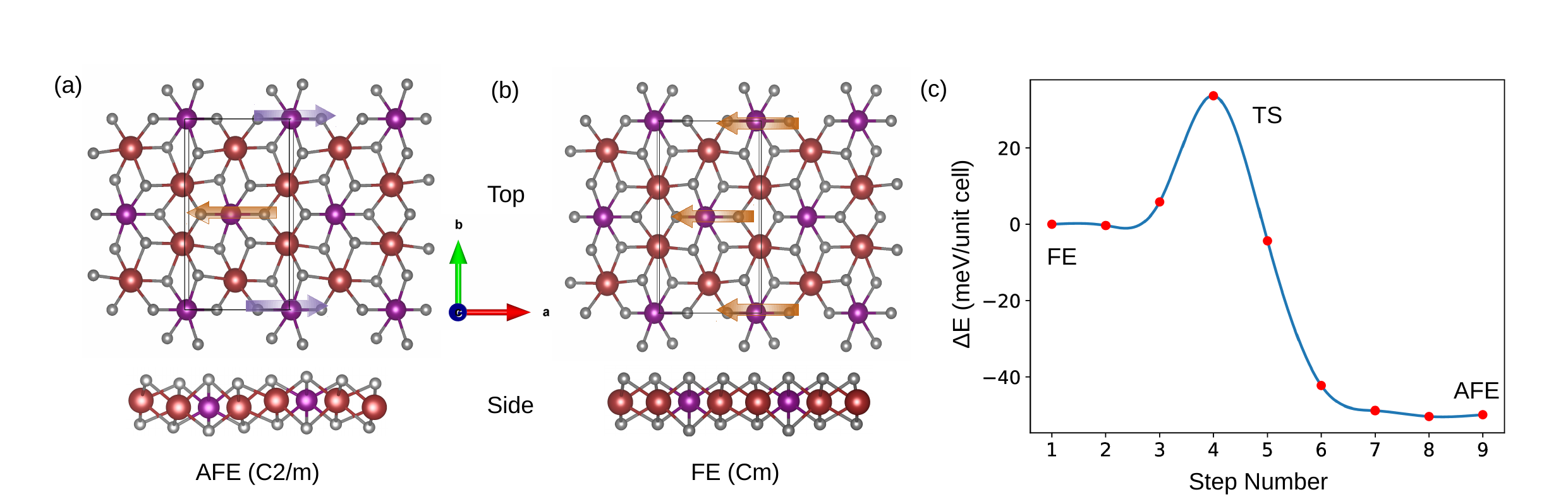}
		\caption{(a) Ferroelectric (FE), (b) Antiferroelectric (AFE) states of Gd-substituted 1T-EuCl$_2$ monolayer (Purple spheres denote Gd atoms), and (c) Minimum energy pathway for transition between FE to AFE state.}  
	\end{figure*}

We performed the first-principles calculations using spin-polarized density functional theory (DFT) as implemented in the Vienna \textit{ab-initio} Simulation Package(VASP) \cite{Kresse}. The core electrons are dealt with using the projector augmented wave (PAW) method. We have used the Perdew-Burke-Ernzerhof (PBE) parametrization \cite{Perdew} of the generalized-gradient approximation as the exchange-correlation functional. The energy cut-off for the plane wave basis was 500 eV. To account for the Hubbard correlation in Gd and Eu 4f-bands, we used the rotationally invariant GGA (PBE) + U method \cite{Liechtenstein} with  U = 9.2 eV and J = 1.2 eV (here U is Coulomb correlation and J is Hund's exchange) for Gd 4f-orbitals and U = 8.5 and J = 1.109 for Eu 4f-orbitals \cite{Gd_nitride}. Treatment of f-electrons using PBE+U is possible as the f-subshell is half-filled. We inserted a vacuum of 23.5\AA between two layers to avoid interlayer interactions. The Hellmann-Feynman forces convergence criterion is set at 0.01 eV/$\AA$ while relaxing the structure, and the energy convergence criterion is $10^{-6}$ eV. A $\Gamma$ centered $15\times11\times1$ Monkhorst-Pack grid was chosen for structural optimization. The phonon dispersion calculations were performed using the Density Functional Perturbation Theory (DFPT) as implemented in VASP and the PHONOPY package with a $2\times2\times1$ supercell \cite{Gonze, Togo}. We used a $7\times5\times1$ k-points grid to construct the Maximally Localized Wannier Functions (MLWFs) and calculate the Berry curvature \cite{wannier1, wannier2}.

In a recent study by Sharan et. al. \cite{sharan}, the rare-earth iodides (RI$_2$) are found to exist in a 2D layered structure in 2H, 1T or 1T$_d$ phases. The ground state phases for different iodides are found to be different. For example, for ScI$_2$, YI$_2$, and GdI$_2$, the ground state is found to be in the 2H-phase, similar to the MoS$_2$ structure, whereas for EuI$_2$, PrI$_2$, and NdI$_2$, the ground state phase is 1T \cite{sharan}. In the 2H-phase, the monolayer of the material has an inherent broken inversion symmetry, thus resulting in inequivalent valleys at the K and K$^{\prime}$ reciprocal lattice points \cite{15,sharan}. Many such materials have been studied extensively in search of applications in the field of Valleytonics. The 1T phase, on the other hand, does not display a broken inversion symmetry in the monolayers, and thus, the valleys are degenerate at the two inequivalent reciprocal lattice points K and K$^{\prime}$. However, it should be noted that even though the 2H-phase monolayers have a broken inversion symmetry, they cannot have a 180$^\circ$ switchable electric polarization because of their C$_3$ symmetry while the 1T-phase has a C$_6$ symmetry and thus can host polarization that can be switched in the opposite direction by means of an external electric field.
	

	In case such an electric polarization is induced in a 1T-monolayer, the inversion symmetry is broken, and a spontaneous valley splitting is simultaneously caused by spin-orbit coupling (SOC) associated with the polarization. The SOC Hamiltonian in the presence of an electric polarization in the system can be written as: 
	
	\begin{equation}\label{first_eq}
	H_{soc} = \lambda(\vec{d}\times \vec{k}).\vec{S} 
	\end{equation}
	where $\lambda$ represents the spin-orbit coupling strength, $\vec{d}$ is the elctric dipole moment, $\vec{k}$ is the crystal momentum and $\vec{S}$ is the spin moment. For the case where the spin moment is locked along out-of-plane (i.e. z) direction and an in-plane electric dipole moment (i.e. \(\vec{d}=(d_x,d_y)\)), the spin Hamiltonian takes the form:
	\begin{equation}\label{second_eq}
	H_{soc}=\lambda_z(d_xk_y-d_yk_x)S_z
	\end{equation}
	
Thus, the SOC energy depends on the crystal momentum and causes a valley-dependent spin-splitting. This splitting also depends upon the direction of dipole moment $\vec{d}$ corresponding to the electric polarization $\vec{P}$. For example, for $d_x \neq 0$, $d_y = 0$, a valley splitting along $k_y$ is obtained, whereas for  $d_y \neq 0$, $d_x = 0$, a valley splitting along $k_x$ is obtained. Furthermore, upon switching the direction of $\vec{d}$ by $180^\circ$, the valley polarization also switches.

In order to realize this mechanism for valley splitting, we begin with the rare-earth halide EuCl$_2$, the iodine counterpart of which has its ground state in the 1T phase according to Sharan et. al. \cite{sharan, note}. In order to confirm that EuCl$_2$ also has its ground state in 1T phase, we performed the first-principles calculations for the 2H and 1T phases of EuCl$_2$. The energy of the 1T phase is found to be 408 meV/formula unit (f.u.) lower than that of the 2H phase. The exfoliation energy for monolayer EuCl$_2$ is found to be 21.34 meV/$\AA^2$. The electronic band structures with SOC for monolayers of 1T and 2H phases of EuCl$_2$ are shown in Fig. S1\cite{sup_mat}. It is clear that the 2H-phase monolayer has a valley splitting of about 47 meV because it does not have an inversion symmetry. On the other hand, there is no valley splitting in 1T-EuCl$_2$ monolayer (ML) because of its inversion symmetry.
	
Next, we perform a 1/3rd Gd-substitution in 1T-EuCl$_2$ ML, where we replace every central Eu-atom of all the Eu-hexarings by a Gd-atom. The formation energy for this substitution is found to be -4.65 eV under Cl-rich conditions, which indicates its feasibility. We calculated the formation energy (E$_f$) using the formula, $$ E_f = E(Eu_2GdCl_6) - E(Eu_3Cl_6) + \mu_{Eu} - \mu_{Gd} $$ where E(Eu$_2$GdCl$_6$) and E(Eu$_3$Cl$_6$) are the energy of Gd-substituted and pristine $ \sqrt{3}\times\sqrt{3}\times1$ supercell of EuCl$_2$ monolayer, respectively, whereas $\mu_{Eu}$ and $\mu_{Gd}$ represent the chemical potentials of Eu and Gd. Under Cl-rich conditions, the chemical potential $\mu_{Eu}$ of Eu is calculated as, $\mu_{Gd} = E(EuCl_2)-E(Cl_2)$.
	
The electronic configuration of Eu in the +2 oxidation state is 4f$^7$5d$^0$, whereas that of Gd is 4f$^7$5d$^1$. Thus, Gd substitution imparts an extra electron to the system. As a consequence of this electron doping, the structure undergoes a bond-centered charge ordering (BCO) in which the extra electron is shared between two Eu and one Gd atoms. This BCO is associated with a shifting of the Gd atoms away from the center of the Eu-hexarings, thus breaking the $C_6$ symmetry of the parent EuCl$_2$ structure. The Gd-shifting can, however, take place in two different ways: (i) The case in which the Gd-atoms in alternate 1D-chains shift along opposite directions. Thus, this structure retains the inversion symmetry of the parent lattice (space group C2/m) and can be termed as an anti-ferroelectric structure (AFE) as shown in Fig 2(a). (ii) The other case involves shifting of all the Gd-atoms along the same (i.e., -x direction as shown in Fig. 2(b)). This structure no longer has an inversion symmetry (space group Cm) and is polar, with the x-axis being the polar axis. The orbital projected density of states (PDOS) for the pristine and Gd-substituted EuCl$_2$ ML in the FE phase are shown in Fig. S2\cite{sup_mat}. We calculated the spontaneous polarization of this polar structure and found it to be 48.40 pC/m. Because of the inversion symmetry of the parent lattice, this polarization can be switched in the opposite direction by means of an external electric field and hence this structure can be referred to as a ferroelectric structure. 
	\begin{figure*}[!t]
		\centering
		\includegraphics[width=15.0cm]{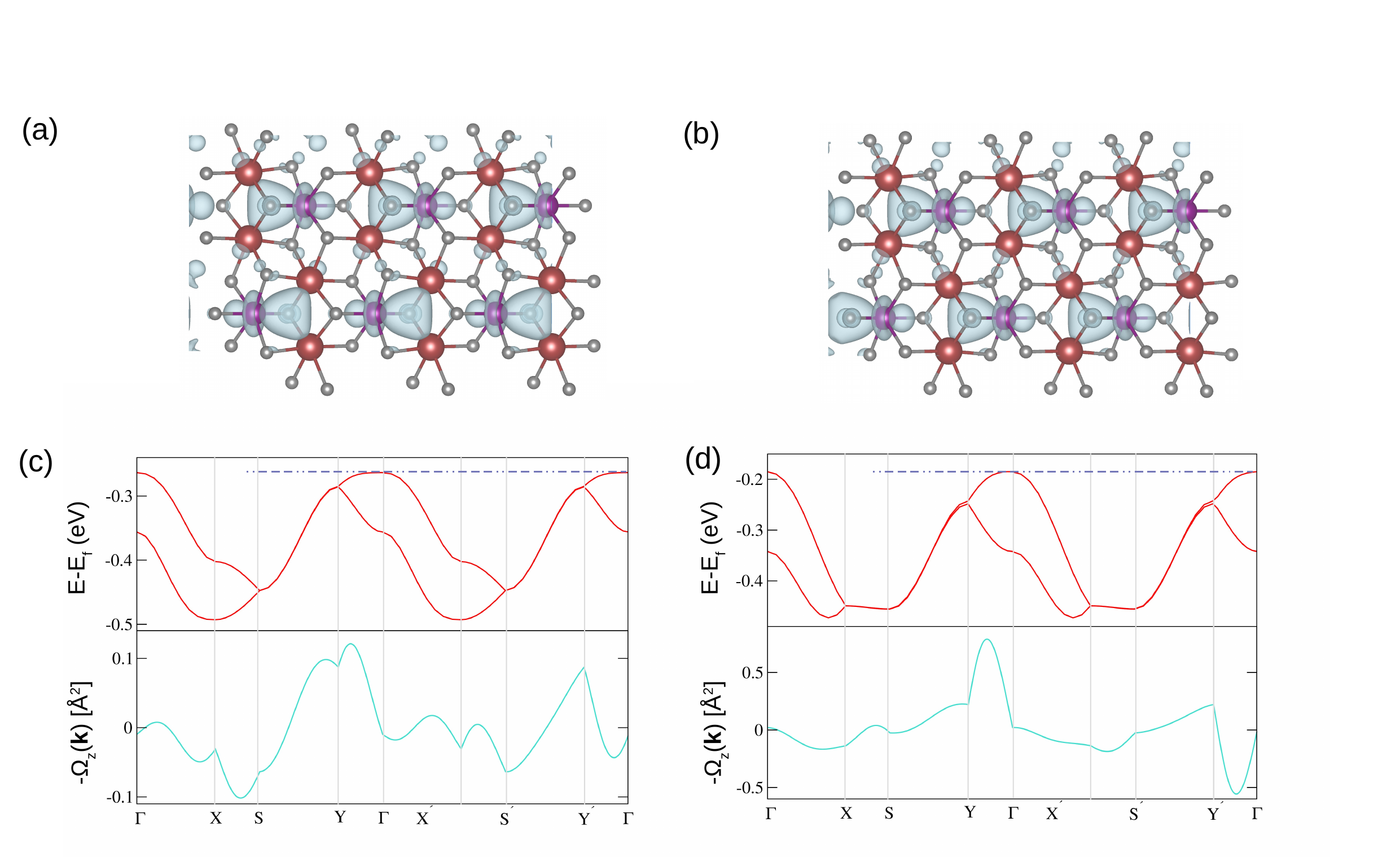}
		\caption{(a) Partial charge density distribution corresponding to the extra electron for FE and (b) for AFE state of Gd-substituted EuCl$_2$ along the monolayer plane consisting of Gd and Eu-ions. (c) Electronic band structure of FE and (d) AFE state along with the Berry curvature with polarization along +x-direction and spin along +z-direction.}
		
	\end{figure*}
	\begin{figure*}[!t]
		\centering
		\includegraphics[width=15.0cm]{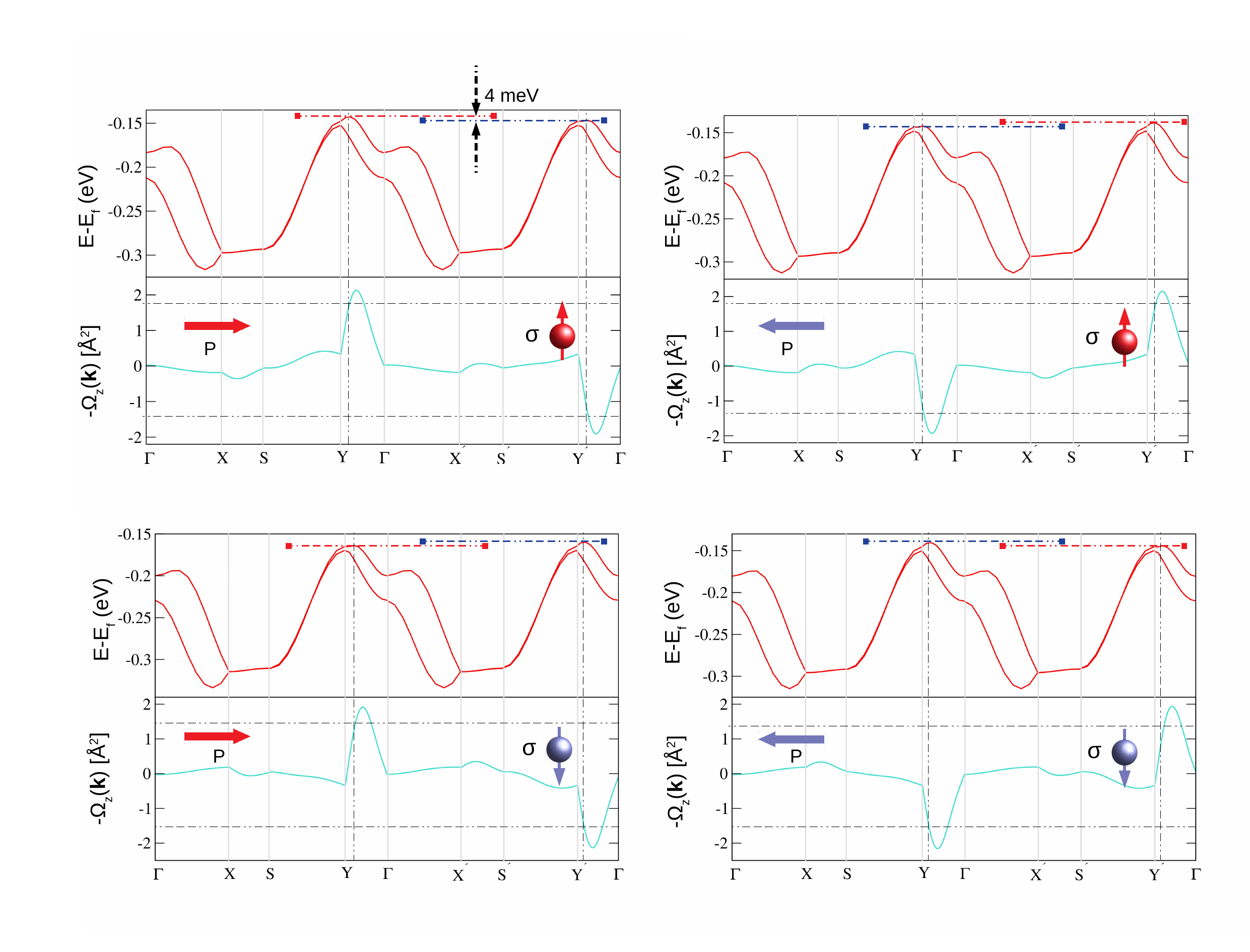}
		\caption{Electronic band structure (with SOC) of monolayer under $5\%$ strain, and Berry curvature, for different polarization and spin directions (A zoomed-in version of the band structure is shown in Fig. S5 \cite{sup_mat}).}
		
	\end{figure*}
	
The nature of BCO in AFE and FE structures can be visualized from the partial charge density plots corresponding to the valence bands, as shown in Figures 3(a) and 3(b), respectively. These plots clearly explain the shifting of Gd-atoms along a single direction in FE and in opposite directions in the AFE structure. The AFE structure is found to have a lower energy than the FE structure by 25.04 meV per formula unit of Eu$_2$GdCl$_6$. However, the FE state can be achieved by applying an external electric field. Both the FE and AFE states are found to be dynamically stable as indicated by their phonon dispersion spectra, where no soft phonons are observed (Fig.S3\cite{sup_mat}). Also, we performed a solid-state nudged elastic band (SS-NEB) calculation to find that the transition from FE to AFE state requires crossing an energy barrier of 16.8 meV per formula unit (Fig.2(c)). This shows that once acquired, the system can stay in the FE phase and will not spontaneously transition into the more stable AFE phase at low temperatures. 
		
Along with the electric polarization, the final structure also possesses a spontaneous strain where the lattice parameter along the y-axis becomes smaller. This is caused by the bond-centered charge ordering, which compresses the bond between Eu-atoms, containing the extra electron from the Gd-atom. The direction of strain and that of polarization are, thus, coupled. Therefore, it should also be possible to switch the direction of polarization by an external mechanical strain \cite{19}. Moreover, as the pristine EuCl$_2$ monolayer has a C$_6$ rotational symmetry, there are a total of six possible directions of electric polarization which should theoretically be achievable by an external electric field or by a mechanical strain.
	
\begin{figure*}[!t]
		\centering
		\includegraphics[width=14.0cm]{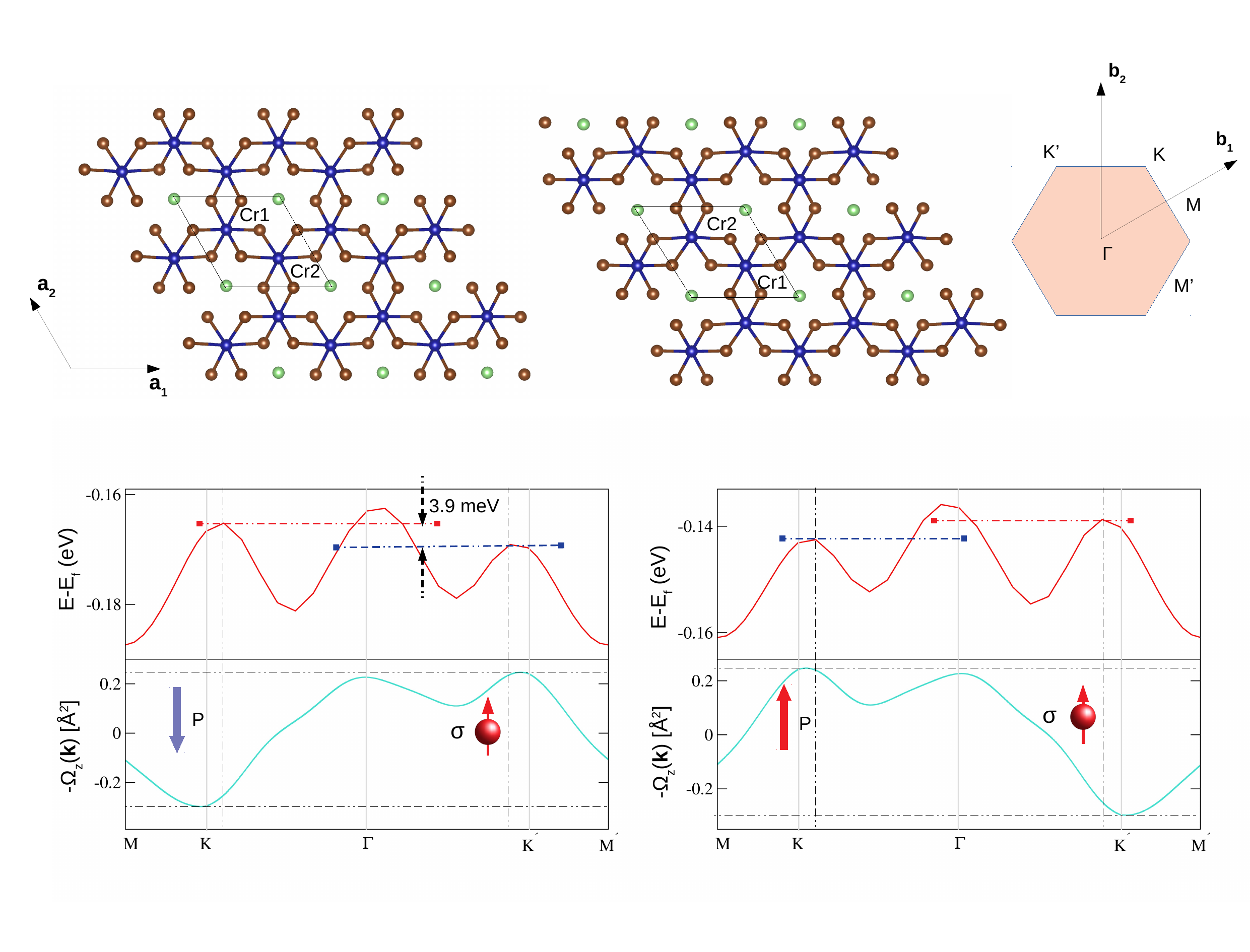}
		\caption{Electronic band structure (with SOC) of (CrBr$_3$)$_2$Li monolayer, and Berry curvature, for different polarization and spin directions.}
		
\end{figure*}

In order to find the magnetic ground state of Eu$_2$GdCl$_6$ monolayer, we calculated the energies of different possible magnetic configurations as shown in Fig. S4\cite{sup_mat}. The energy of ferromagnetic (FM) configuration is found to be 70.33 meV/f.u. and 9.13 meV/f.u. lower than that of the AFM1 and AFM2 magnetic configurations (Fig. S4). Thus, the ground state magnetic configuration of the monolayer is FM. To find the spin easy-axis, we calculated the magnetic anisotropy energy defined by $E_c-E_{ab}$, where $E_c$ represents the energy of the monolayer with spins oriented out-of-plane of the monolayer and $E_{ab}$ denotes the energy when the spins are aligned in-plane. The MAE for spins oriented parallel or anti-parallel to the polar axis is 1.206 meV/f.u whereas the MAE for spins pointing perpendicular to the polar axis is -168 $\mu$eV/f.u. Thus, the spin easy-axis is found to be perpendicular to the polar axis. Hence, the magnetization and polarization axes are coupled in the ferroelectric (FE) Eu$_2$GdCl$_2$ monolayer. This opens up the possibility for electric control of magnetization due to FM-FE coupling in this material, a highly sought-after property from the aspect of less energy consumption. However, it should be noted that the spin easy-axis, which is in a direction perpendicular to the polar axis, does not change direction when the polarization is switched by 180$^\circ$. Therefore, upon reversing the polarization with an electric field, the spin orientation remains the same. But if the polarization axis is changed by an angle other than 180$^\circ$, the spin easy-axis also rotates so as to again become perpendicular to the new polar axis.	

In addition to the FM-FE coupling, we also expect another important property in the FE monolayer, i.e., the ferroelectric-ferrovalley (FE-FV) coupling. To check this, we examine the electronic band structure of the monolayer in both AFE and FE phases with spin-orbit coupling (SOC) at play. Now, the AFE state still has the inversion symmetry, and the net electric dipole moment per unit cell $\vec{d}$ is zero. Therefore, the AFE state can not exhibit any valley splitting due to spin-orbit coupling (SOC), which can be seen in Fig. 3(c). The FE state, on the other hand, no longer has the inversion symmetry of the parent EuCl$_2$ monolayer (ML) and has a finite electric dipole moment $\vec{d}$. Therefore, it has a non-zero SOC corresponding to the equation (\ref{second_eq}), which is crystal momentum dependent. The electronic band structure with SOC confirms this as energy eigenvalues of the top two valence bands are found to be unequal along the paths Y-$\Gamma$ and Y$^{\prime}$-$\Gamma$ (Fig. 3(d)). We have also shown the corresponding Berry curvatures for AFE and FE states in Figures 3(c) and 3(d). The valley, i.e., the valence band maximum (VBM), however, lies on the $\Gamma$ point and, hence, no valley splitting is observed for the FE structure either. This problem is resolved by applying a biaxial tensile strain of $5\%$ on the monolayer under which the VBM shifts away from the $\Gamma$ point and toward the Y/Y$^{\prime}$-point, as is shown in Fig. 4. Thus, the VBM now lies between Y/Y$^{\prime}$ and $\Gamma$. The Berry curvature peaks are also obtained between Y/Y$^{\prime}$ and $\Gamma$. The Berry curvature peak value along Y-$\Gamma$ is found to be \(-2.12 \ \AA^2\), whereas that along Y$^{\prime}$-$\Gamma$ is \(1.96 \  \AA^2\).
		
	As per the equation (\ref{second_eq}), the valley degree of freedom is directly coupled to the direction of polarization, and thus, upon switching the polarization direction, one should be able to achieve valley switching as well. This is confirmed by the electronic band structure (with SOC) of the monolayer with the electric polarization along the x direction shown in Fig. 4. It can be seen in Fig. 4(a) that for polarization along x direction, the VBM corresponds to the path Y-$\Gamma$ whereas for the polarization along +x direction, the VBM lies between Y$^{\prime}$-$\Gamma$. The values of Berry curvature along Y-$\Gamma$ and Y$^{\prime}$-$\Gamma$ are also interchanged. The value of valley splitting is found to be about 4 meV, which is comparable to the values found in kagome lattice Ti$_3$B$_8$ \cite{7} and the bilayer 2H-VSe$_2$ \cite{11}. It is to be noted that the valley splitting along the path Y/Y$^{\prime}$-$\Gamma$ is consistent with the expression for SOC given above in equation (\ref{second_eq}) for an electric polarization along the x-direction. 
	
Thus, we demonstrate that because of the coupling between the electric and valley polarization in the Gd substituted EuCl$_2$ monolayer, the valley polarization can be switched by an external electric field, unlike in the conventional ferrovalley materials, in which an external magnetic field is required to switch the valley degree of freedom. Besides, the valley splitting can also be switched in the conventional way, i.e., by changing the spin orientation by using an external magnetic field. This can be seen in the electronic band structures shown in Fig. 4(c) and 4(d). It should be noted, however, that in order to have a finite splitting due to SOC, the spins should align along the z-axis (equation (\ref{second_eq})).  Therefore, we still need to apply an external magnetic field to keep the spin orientation fixed along the z-axis. However, one does not need a changing magnetic field to switch the valley degree of freedom. Instead, by switching electric polarization, one can directly obtain valley switching.

	We also studied the valley splitting in another such material, which exhibits in-plane polarization, i.e., Li- intercalated CrBr$_3$ monolayer \cite{crbr3_prl}. This material involves a site-centered charge ordering (SCO) along with an orbital ordering (OO) due to the extra electron imparted by Li-intercalation. This system undergoes an asymmetric Jahn-Teller (JT) distortion at alternate Cr-sites. This combination of charge and orbital ordering gives the material a polar symmetry. We, therefore, expected a similar valley splitting in (CrBr$_3$)$_2$Li as well. To confirm this, we calculated the electronic band structure of (CrBr$_3$)$_2$Li monolayer with SOC and found that the valleys at K and K$^{\prime}$ have a splitting of about 3.9 meV. Here, also, the SOC depends upon the direction of in-plane polarization and the valley degree of freedom can be switched by changing the direction of electric polarization (Fig.5). Thus, the ferroelectricity-ferrovalley (FE-FV) coupling provides an electric field control of the valley in (CrBr$_3$)$_2$Li monolayer as well. The spin-easy axis, in this case, lies in the plane of the monolayer parallel or anti-parallel to the direction of the polar axis \cite{crbr3_prl}. Thus, in order to observe the valley splitting due to SOC, one must apply an external magnetic field to bring the spin easy-axis out of plane of the monolayer, in this case as well.
	
In summary, we propose a novel route to induce ferroelectric-ferrovalley coupling where valley polarization can be switched by an external electric field. We demonstrate using first principles density functional theory calculations considering two real materials (a rare-earth metal halide i.e. Gd-substituted EuCl$_2$ monolayer and a transition metal halide i.e. Li-intercalated CrBr$_3$ monolayer) the feasibility of this phenomenon. We have shown that the in-plane electric polarization induced by the bond or site centered charge ordering in these systems drives the spontaneous valley splitting via spin-orbit coupling (SOC) associated with the electric polarization.  	
\\	
	
	\noindent SB acknowledges Council of Scientific and Industrial Research (CSIR), India for research fellowship. TM acknowledges Science and Engineering Research Board (SERB), India for funding support through MATRICS research grant (MTR/2020/000419). TM and SB acknowledge the National Supercomputing Mission (NSM) for providing computing resources of ‘PARAM Ganga’ at the Indian Institute of Technology Roorkee, which is implemented by C-DAC and supported by the Ministry of Electronics and Information Technology (MeitY) and Department of Science and Technology (DST), Government of India.


\begin{thebibliography}{99}
\bibitem{note} We note that Gd substitution in EuI$_2$ doesn't give rise to an electric polarization as we demonstrate for EuCl$_2$ here.
\end{thebibliography}
\end{document}